\documentclass[twocolumn,showpacs,preprintnumbers,amsmath,amssymb]{revtex4}
\usepackage{graphicx}
\usepackage{amsmath}
\usepackage{amssymb}
\usepackage{bm}
\usepackage{epstopdf}

\begin{document}
\textbf{Comment on "Fock-Darwin States of Dirac Electrons in
Graphene-Based Artificial Atoms"}

The authors of Ref.\cite{Chen} have computed eigenenergies and
eigenstates of a graphene dot in the presence of two-dimensional
(2d) harmonic potential and a magnetic field by considering only
basis states with energies larger than or equal to zero.  We show
that the wavefunctions  of the nearly chiral Landau level (LL)
obtained using this approximate method are inaccurate, and  violate
a Hellman-Feynman theorem.  Corrections lead to
significant changes of optical strengths involving nearly chiral
states. Moreover, additional basis states lead to new energy levels
in the energy spectrum.

\begin{figure}[!hbpt]
\begin{center}
\includegraphics[width=0.4\textwidth]{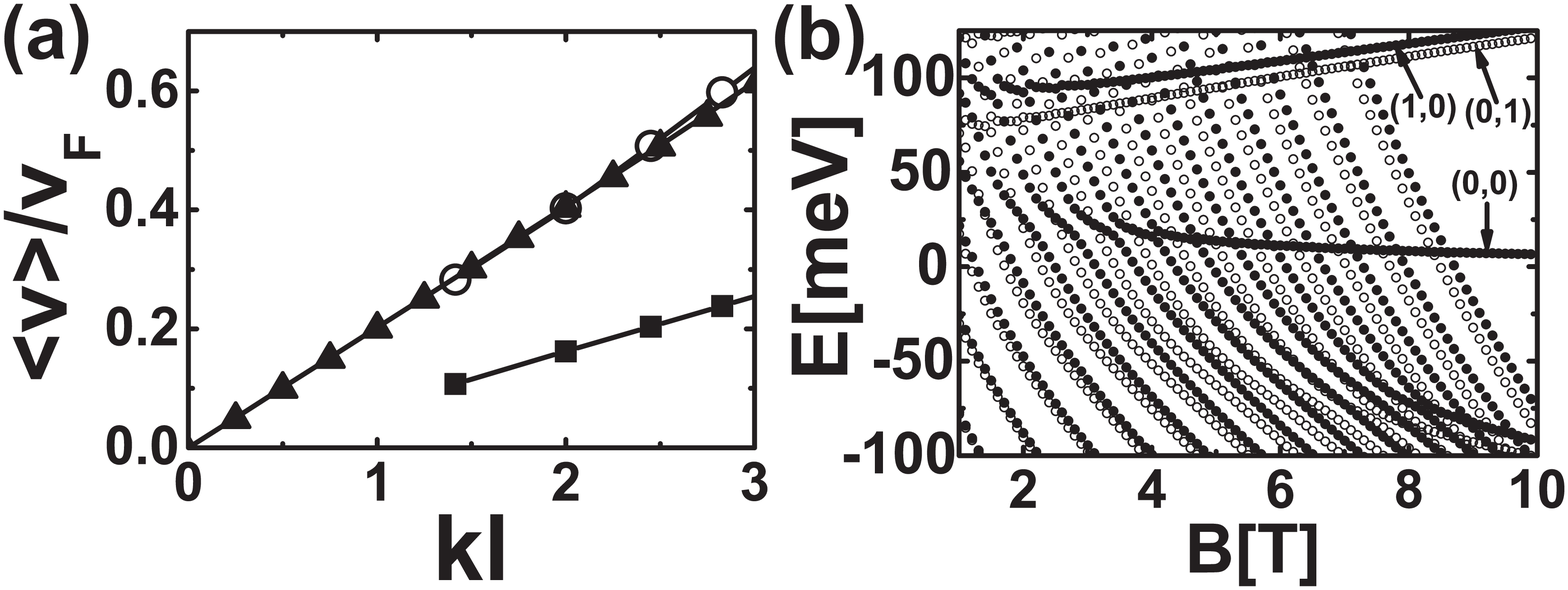}
\caption{
%(a) Velocity
%$\vec{v}(\vec{r})$ of the eigenstate with $n=0$ and $m=0$ $(k\ell=\sqrt{2})$ when both %positive
%and negative energy basis states are included.
%(b)
%$\vec{v}(\vec{r})$ when negative energy basis states are excluded.
(a) Velocity $\langle v\rangle$ of the eigenstate with $n=0$.  Here
$\alpha=(\kappa \ell^2)/(\frac{\hbar v_F}{\ell})=0.2$.  Values of
velocities using $21$ basis states: $\frac{\langle v_{\theta}
\rangle}{v_{F}}$ is for 2d harmonic potential (circles), and
$\frac{1}{\hbar}\frac{\partial \epsilon_0(k)}{\partial
k}\frac{1}{v_{F}}$ are for 1d and 2d harmonic potential systems
(triangles).  $\frac{\langle v_{\theta}\rangle}{v_{F}}$ is for 2d
harmonic potential computed excluding negative energy basis states
(squares).  $11$ basis states are included.  (b)
%Hilbert space is divided into subspaces with different angular momentum $J$\cite{Park} and a %Hamiltonian matrix is solved in each subspace.
Energy levels  in the Hilbert subspaces with angular momentum
$J=-1/2$ (filled  circles)  and $J=1/2$ (open circles) are shown
together. The number basis states used is  $30$.   Allowed  optical
transitions obey $\Delta J=\pm1$.
%Energies of the lines denoted by  $(0,0),  (0,1)$, and  $(1,0)$ agree with the corresponding  %results  in Fig.3 of Chen et al\cite{Chen}, except in interval $B<4$T where  anticrossings of %$(0,0)$ with other states take place.
%  (Note that definitions of $(n,m)$ are different in our work and theirs).
% Here  numbers of  basis states in a Hilbert subspace of $J$ are different between Chen et al's %appraoch and ours: they use  $15$ positive energy basis states and we use $30$ positive and %negative basis states.
}\label{fig:graph}
\end{center}
\end{figure}

In Ref.\cite{Chen} the following type of basis states of the Hamiltonian
matrix  are used with energies larger than or equal to zero
\begin{eqnarray}
\psi_{n,m}(\vec{r})=D_{n}\left(\begin{array}{c} -\textrm{sgn}(n)i\phi_{|n|-1,m}(\vec{r})\\
\phi_{|n|,m}(\vec{r})\end{array}\right). \label{wavef}
\end{eqnarray}
The energies of these states are $E_n=\textrm{sgn}(n)\frac{\hbar
v_F}{\ell}\sqrt{2|n|}$ ($v_F$ is the Dirac velocity and $\ell$ is
the magnetic length).
%First we show that the value of azimuthal velocity for the nearly
%chiral LL ($n=0$) states can be computed correctly only by including
%both positive and negative energy basis states.
%To demonstrate our
%point in a simple way we apply
First order perturbation theory  gives the perturbed wavefunctions of nearly chiral
LL states with $n=0$:
$|\psi^{(1)}_{0,m}\rangle\approx
|\psi_{0,m}\rangle+C_{1}|\psi_{1,m+1}\rangle+C_{-1}|\psi_{-1,m+1}\rangle$.
%Note that LL states $|\psi_{0,m}\rangle$ are chiral, i.e., only
%their second component in Eq.(\ref{wavef}) is non-zero, but the
%other terms $|\psi_{1,m+1}\rangle$ and $|\psi_{-1,m+1}\rangle$ are
%non-chiral.
These states have all the same angular momentum
$J=-m-1/2$.
%The expansion coefficients are $C_{\pm 1}=\frac{\langle
%\psi_{\pm 1,m+1}|V|\psi_{0,m}\rangle}{-E_{\pm 1}}$, where 2d
%harmonic potential is $V(r)=\frac{1}{2}\kappa r^2$.
We find, using $C_{-1}=-C_{1}$,  that the expectation value of the azimuthal velocity
$\langle v_{\theta}\rangle$ is {\it two times} bigger when both
positive and negative energy basis states are included than when
only positive energy basis states are included in the perturbation
series:
%\begin{eqnarray}
$\langle \psi^{(1)}_{0,m}|v_{\theta}|\psi^{(1)}_{0,m}\rangle
 =2\sqrt{2}v_{F}Re\Big[C_{1}\langle
\phi_{0,m}|e^{i\theta}|\phi_{0,m+1}\rangle\Big]$.
%\end{eqnarray}
Our diagonalization of the Hamiltonian matrix confirms the simple
perturbative result above.
%The results in Figs.\ref{fig:graph}(a)
%and (b) demonstrate that there is a significant difference in
%$\vec{v}(\vec{r})$ of a $n=0$ state when both positive and negative
%energy basis states are included and when only positive energy basis
%states are included.
Fig.\ref{fig:graph}(a) shows that the
difference in $\langle v_{\theta}\rangle$ is about a factor of two.
%Ref.\cite{Gia}  explains that  the hermiticity of the Hamiltonian
%matrix is consistent with deconfined behavior of two component Dirac
%wavefunctions.

The azimuthal velocity $\langle v_{\theta}\rangle$ of states in a
quantum dot with $V(r)=\frac{1}{2}\kappa r^2$  should approach the
velocity along a Hall bar\cite{Yang} in the 1d harmonic potential
$\frac{1}{2}\kappa x^2$. It is easy to show that the graphene edge
electrons of a Hall bar obey the  Hellman-Feynman theorem
$\frac{1}{\hbar}\frac{\partial \epsilon_n(k)}{\partial
k}=\langle\psi_{n,k}|v_y|\psi_{n,k}\rangle$, where the y-component
of wavevector is denoted by $k$ and the energy of the $n$'th LL by
$\epsilon_n(k)$. Fig.\ref{fig:graph}(a) displays the values of
$\langle v_{\theta}\rangle$ for nearly chiral states of a dot, and
we see that they converge to $\frac{1}{\hbar}\frac{\partial
\epsilon_0(k)}{\partial k}$ of the Hall bar (the relevant quantum
numbers of the two systems are related by
$k\ell=\sqrt{2(m+1)}$\cite{Yang}).  When negative energy basis
states are excluded in the Hamiltonian matrix the value of the azimuthal velocity $\langle
v_{\theta}\rangle$  is reduced
significantly  while the value
$\frac{1}{\hbar}\frac{\partial \epsilon_0(k)}{\partial k}$ does not
change as much.  The Hellman-Feynman theorem is thus violated in
this case.

%Inaccurate wavefunctions of nearly chiral states also
%lead to significant deviations of the strength of the optical
%transitions reported in Ref.\cite{Chen}.  For example,

Moreover,  numerous new states appear in the energy spectrum, for
example, between the energies of $(0,0)$ and $(0,1)$, as shown in
Fig.\ref{fig:graph}(b).  Note also there are {\it negative} energy
levels, which are absent in Ref.\cite{Chen}.  Some of these new
levels  give rise to additional optical transitions with significant
strength. The appearance of additional levels with positive energies
is due to  negative energy basis states that get perturbed by the
parabolic potential with their energies shifted upward.  In
addition, we find that the strength of the optical transition
$(0,0)\rightarrow (0,1)$ at B=$2$T with $\alpha=0.925$ and
polarization along x-axis, shown in Fig.3 of Chen et al.\cite{Chen},
is about  $2.6$ times larger than the correct value obtained by
including  both positive and negative energy basis states. Physical
origin of this difference  is anticrossing of $(0,0)$ with other
levels in the interval $B<4$T, see Fig.\ref{fig:graph}(b).  The
states $(1,0)$ and $(0,1)$ also anticross below $2$T with new levels
arising from the presence of negative energy basis states.  This
effect is missing in Ref.\cite{Chen}, and leads to additional
modifications of optical strength.

%The Hamiltonian used in Ref.\cite{Chen} has valley symmetry.  Adding a mass term breaking %valley symmetry\cite{Rec} would give an improved model Hamiltonian of a parabolic dot.

Acknowledgment. This work was supported by the Korean Research
Foundation Grant funded by the Korean Government (KRF-2009-0074470).
\\\\
\noindent
P.S. Park, S.C. Kim, and S. -R. Eric Yang$^{*}$\\
%\footnote{corresponding author,eyang812@gmail.com}}\\
Physics Department, Korea University,
Seoul Korea\\
*eyang812@gmail.com
\\

\end{document}